\documentstyle[11pt,paspconf,epsf]{article}

\begin{document}

\title{Rapid variability of BL Lacs and time lags:\\
what can we learn?}
\author{M. Chiaberge, A. Celotti}
\affil{S.I.S.S.A.,via Beirut 4, I-34014 Trieste, Italy}

\author{G. Ghisellini}
\affil{Osservatorio Astronomico di Brera, V. Bianchi 46, I-22055 Merate, Italy}




\begin{abstract}

The quasi--symmetric shape of the lightcurves, as seen in the 
X--rays and in the optical, 
together with the fast variations of the flux observed, suggest 
that the cooling time for the highest energy electrons may be 
shorter than the light crossing time $R/c$.
Moreover the existence of time lags between lightcurves at 
different frequencies can be interpreted, in the frame of 
homogeneous models, as due to 
the different cooling times of electrons of different energy.
To reproduce in detail the variability pattern at different 
wavelengths we need to study the time dependent behavior of the
emitting particle distribution and to take into account of the different light 
travel times of photons produced in different regions of the source.
We apply the model to PKS 2155--304 (1991 and 1994 campaigns).

\end{abstract}

\section{The model}
The source, of typical dimension $R$, is embedded
in a tangled magnetic field $B$.
Relativistic electrons are injected homogeneously
throughout the source for a time $R/c$: this is equivalent to
the case in which a shock active for a time $R/c$ runs along a 
region of the jet (Chiaberge \& Ghisellini 1998).
We consider Synchrotron and Synchrotron Self 
Compton cooling, and particle escape (assumed being independent 
of energy). 
The source moves relativistically with a Lorentz 
bulk factor $\Gamma$, and the radiation is beamed with a Doppler 
factor $\delta$.
In order to reproduce the fast variability, 
a flaring emission is summed to a constant component.
We assume that the source is 
a cube of dimension $R$, 
moving towards the observer at an angle $\theta=1/\Gamma$ to the line of
sight, which is appropriated for blazars. 
In the comoving frame, the transformed angle is $\theta^{\prime}
=90^{\circ}$.
We numerically solve the continuity equation for the electron 
distribution and we calculate the SSC spectra produced at any 
timestep. We consistently take into account, as already mentioned,
the different light travel times of photons produced in different 
regions of the source (for details see Chiaberge \& Ghisellini, 1998).
Notice that in the case of a ''cubic`` geometry the different slices have 
equal volumes: in this way we can separate the geometrical effects
from the radiative cooling effects. Extension to different 
geometries (e.g. cylinder, sphere) can be trivially taken into account
by properly weighting the different slice volumes.

\subsection{Application to PKS 2155-304}

We consider both the 1991 (Edelson et al. 1995) and the 1994 
(Urry et al. 1997) multiwavelength campaigns. 
Time delays between lightcurves at different frequencies and 
quasi--symmetric flares (especially in the X--ray band)
were observed.
Nevertheless the two campaigns showed different behaviors
with respect to the time lags between the X--rays and the lower 
frequencies.
We qualitatively reproduce the behavior observed 
during the 1991 campaign, where the time lags between the X--ray
band and the UV/optical was found to be $\sim 3$ hours.
The 1994 X--ray flare shape 
can be well reproduced by our model with almost the same physical 
parameters used for the 1991 case, except for a different injection function.
In fact, a power law injection, appropriate to reproduce the 1991 data,
would cause the UV and optical fluxes to rise 
almost simultaneously to the X--ray flare, contrary to what observed. 
A ''monoenergetic`` injection function can reproduce this
particular behavior.
A fundamental problem remains in explaining the data: in order
to reproduce symmetric X--ray and UV flares we require that the electron 
cooling time is lower than $R/c$ (which is constrained by the flare 
duration) in a wide range of energies. A short $t_{cool}$ requires an intense 
magnetic field, but 
this immediately implies that also the time lag between X--rays and UV
is too short to agree with the observed 3--4 days delay.
Two alternative possibilities can be envisaged:
(i) introduce some inhomogeneities in the model;
(ii) the flare revealed by ASCA is not associated to 
the UV one, and the correlation between the two is artificially due to
the lack of data in the X--ray band.

\begin{figure}
\plottwo{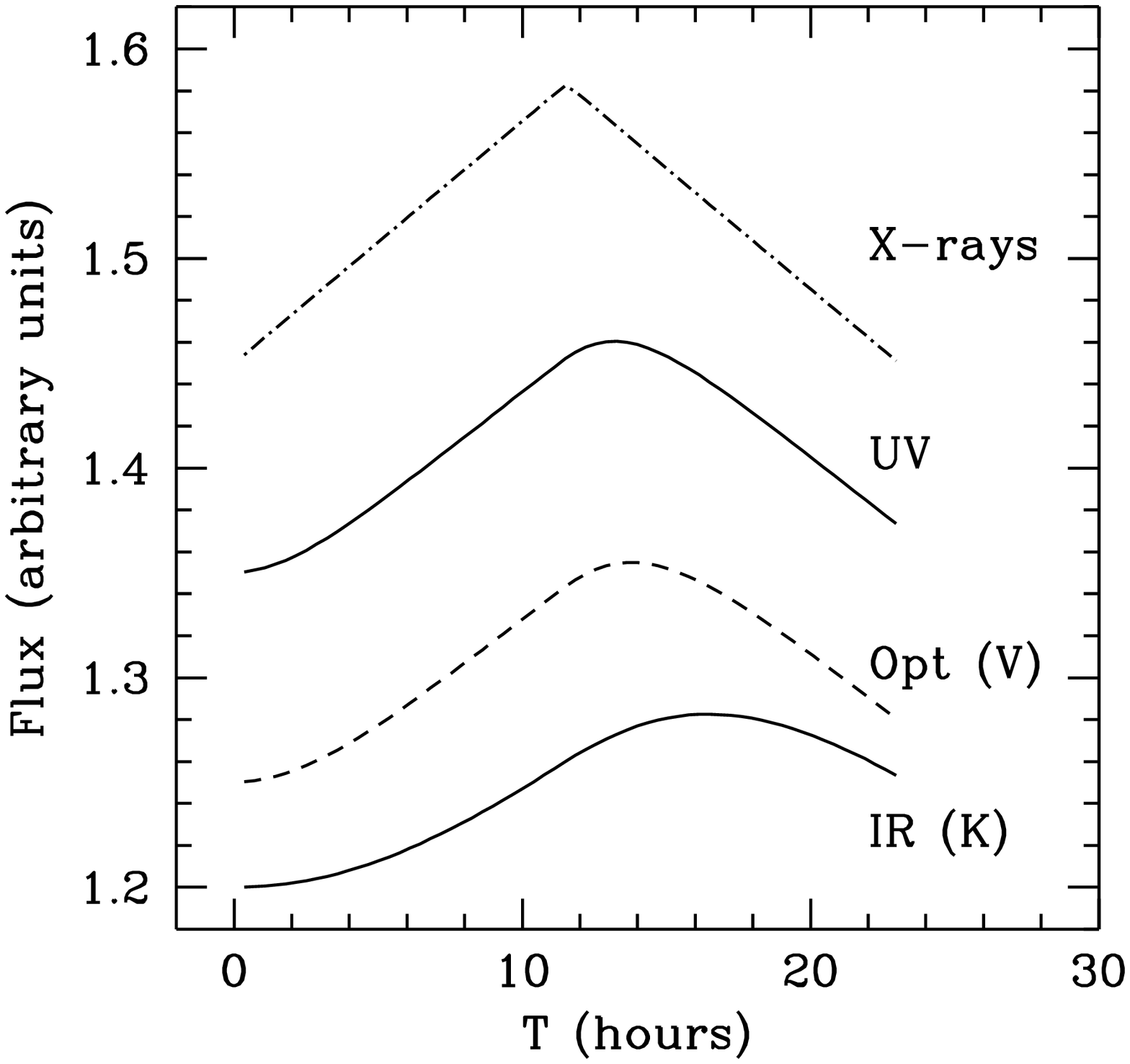}{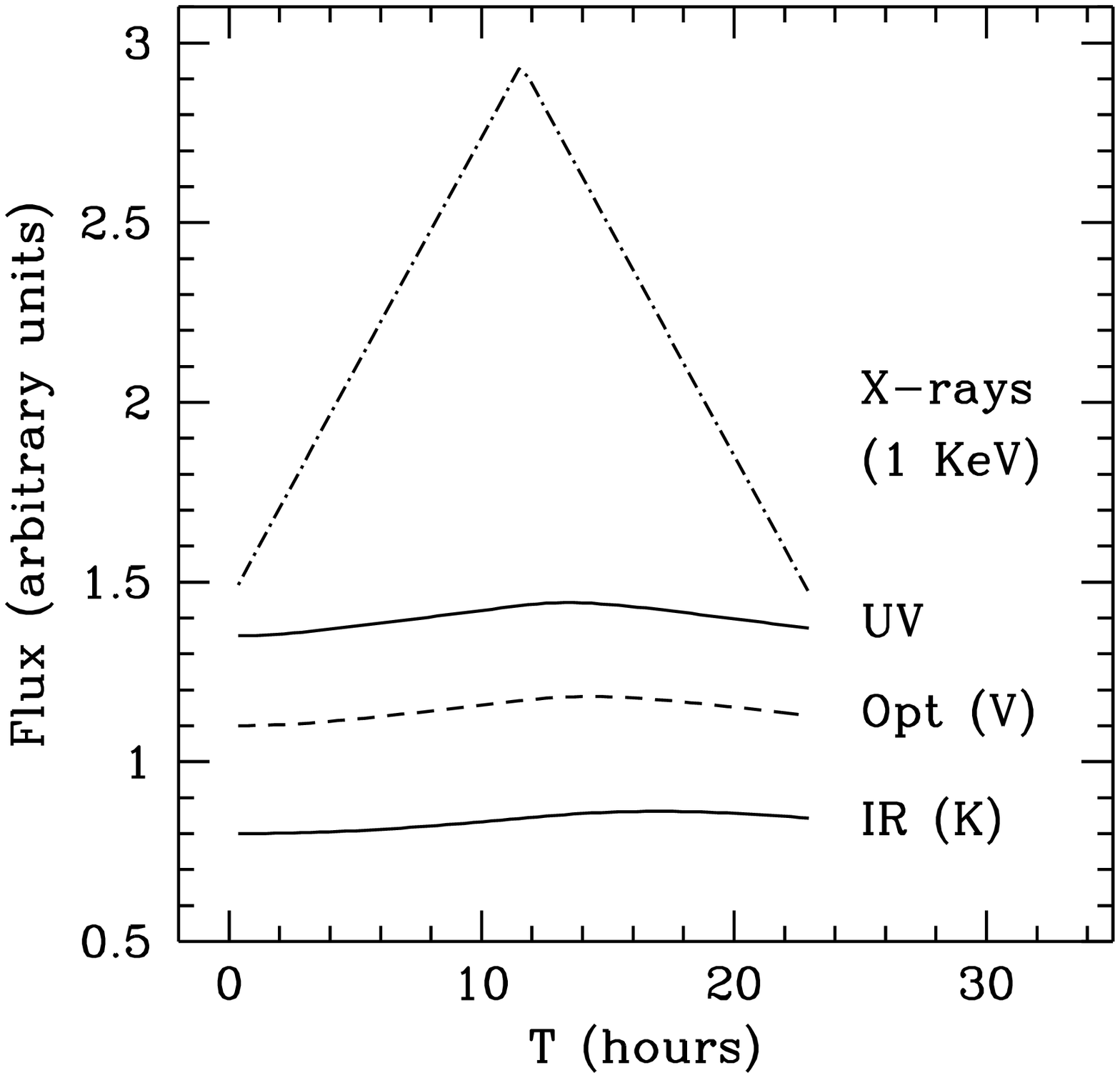}
\vspace{-0.8cm}
\caption{\footnotesize\sf\protect\baselineskip 8pt 
Simulated lightcurves for the 1991 (left) and 1994 (right) campaigns.
Physical parameters: $R=2 \times 10^{16}$ cm, $B=1$ G, $\delta=18$,
$L_{inj}=1.5 \times 10^{41}$ erg/s (1991);
$L_{inj}=5.3 \times 10^{41}$ erg/s (1994).
Injection laws:
$Q(\gamma)\propto \gamma^{-2.8}$, $\gamma_{min}=1.2 \times 10^{4}$,
$\gamma_{max}=10^{5}$, $t_{inj}=R/c$ (1991);
narrow gaussian $\gamma_{peak} =10^{5}$,  $t_{inj}=R/c$ (1994).}
\vspace{-0.6cm}
\label{figura}
\end{figure}

\end{document}